\documentclass[authoryear]{elsarticle}

\usepackage{natbib}

\usepackage{graphics}
\usepackage{graphicx}
\usepackage{epsfig}
\usepackage{color}           

\usepackage{amssymb}

\newcommand{\aap}{{\it Astron. Astrophys. }}

\newcommand{\apj}{{\it Astrophys. J. }}
\newcommand{\apjl}{{\it Astrophys. J. Lett. }}

\newcommand{\mnras}{{\it Mon. Not. Roy. Astron. Soc. }}

\newcommand{\pasj}{{\it Pub. Astron. Soc. Japan }}

\newcommand{\solphys}{{\it Solar Phys. }}

\newcommand{\ssr}{{\it Space Sci. Rev. }}

\begin{document}

\begin{frontmatter}



\title{Study of multi-periodic coronal pulsations during 
an X-class solar flare}


\author{Partha Chowdhury$^a$, A.K. Srivastava$^b$, B.N. Dwivedi$^b$, Robert Sych$^{c}$, Y.-J. Moon$^{a}$}

\address{$^a$School of Space Research, Kyung Hee University, Yongin, Gyeonggi-Do, 446-701, Korea.}
\ead{partha@khu.ac.kr,  moonyj@khu.ac.kr}
\address{$^b$Department of Physics, Indian Institute of Technology (Banaras Hindu University), Varanasi-221005, India.}
\ead{asrivastava.app@iitbhu.ac.in, bnd.app@iitbhu.ac.in,}
\address{$^c$Institute of Solar-Terrestrial Physics SB RAS, Irkutsk-664033, Russia.}
\ead{sych@iszf.irk.ru}

\begin{abstract}
We investigate quasi-periodic coronal pulsations during the decay phase of an X 3.2 class flare on 14 May 2013, using soft X-ray data from the RHESSI satellite. Periodogram analyses of soft X-ray light curves show that $\sim$ 53 s and $\sim$ 72 s periods co-exist in the 3--6, 6--12 and 12--25 KeV energy bands. Considering the typical length of the flaring loop system and observed periodicities, we find that they are associated with multiple (first two harmonics) of fast magnetoacoustic sausage waves. The phase relationship of soft X-ray emissions in different energy bands using cross-correlation technique show that these modes are standing in nature as we do not find the phase lag.  Considering the period ratio, we diagnose the local plasma conditions of the flaring region by invoking MHD seismology. The period ratio ~P$_{1}$/2P$_{2}$ is found to be $\sim$ 0.65, which indicates that such oscillations are most likely excited in longitudinal density stratified loops.
\end{abstract}


\end{frontmatter}

\clearpage
\section{Introduction}

Solar flares are sudden release of magnetic energy occurring in the solar atmosphere, 
and lasting from a few tens of seconds (impulsive) to a few tens of minutes (gradual). 
Flare emissions are detected in the entire electromagnetic spectrum, ranging 
from radio, microwave, visible, ultraviolet, X-rays, hard X-rays, and even to Gamma-rays. The effect of these flares 
are also detected in the variations of the flux density of solar energetic particles 
approaching towards the Earth. The electromagnetic radiation generated in solar flares 
often exhibits oscillatory patterns in the light curves. They are periodic in nature, 
operating in the flare magneto-plasma, and having typical periods ranging from a few 
milliseconds to several minutes \citep[and references cited therein]{Nak2005,Nak2009,Kim2012,Kup2013,Huang2014}. 
These oscillations are referred to as quasi-periodic pulsations (QPPs). In some cases, 
they have inharmonic shape with apparent amplitude and period modulation \citep{Nak2009}.


QPPs are classified mainly as short QPPs (sub-second), medium QPPs (seconds to several minutes), 
and long QPPs (from several minutes to tens of minutes) according to the possible physical mechanisms 
associated with them \citep{Nak2009,Kup2010}. According to \cite{Asc1987}, the short QPPs are probably 
connected with the interaction of electromagnetic, plasma, or whistler waves with accelerated particles 
detected in the radio emission. Medium QPPs are likely associated with magnetohydrodynamic (MHD) 
processes in the solar flaring loops as detected in radio, microwave, white light, and X-ray emissions 
\citep[and references cited therein]{Ing2008,Ing2009,Jak2008,Asai2001,Mel2005,Rez2007,Mc2005,Rez2011}. 
On the other hand, the long ones are usually relevant to active region (AR) dynamics \citep{Rob2009,Rob2015}
and global oscillations of the Sun \citep{Tan2010,Srivastava2010}.

QPPs in solar flares are one of the major diagnostic tools to study the physical conditions 
in flaring sites, and triggering mechanisms. These studies can also be extended to stellar 
flares, which exhibit QPPs in their radio, optical, and soft X-ray emissions 
\citep{Math2003,Mitra2005,Pandey2009,Anfinogentov2013,Srivastava2013}. 
Understanding QPPs during the flare eruption is still an open question. 
The fundamental physical processes of these oscillations in solar flares 
are relevant to understanding magnetic reconnection, magnetohydrodynamic 
(MHD) waves in coronal structures, particle acceleration, thermodynamics, 
and other kinetic effects.  Although the properties of QPPs in solar hard
 X-ray (HXR) emissions have been widely studied by different researchers, their nature in solar soft X-ray emissions (SXRs) in different energy 
bands simultaneously remains to be extensively investigated. This calls 
for more observational information from samples with oscillatory behavior 
in low-energy X-rays.

The ratio of the statistically significant QPPs of the simultaneously 
existing spectral components is important for determining the nature and 
mechanisms of the pulsations both in observations and theory 
\citep[and references cited there]{Tom2007,Srivastava2008,Inglis2009,
Srivastava2010,Mac2011,Orza2013,Luna2012,Robert2014,Yang2015}. 
Different periods of QPPs may be associated 
either with different MHD modes simultaneously present in a flaring loop or 
with different harmonics of the same MHD mode \citep[e.g.]{Mel2005, 
Tom2007,Srivastava2008,Andries2009,Inglis2009,Srivastava2010}. 

In this paper, we use the RHESSI data to study the intensity of the 
co-existing QPPs at 3--6 KeV, 6--12 KeV and 12--25 KeV multiple energy 
bands during the decay phase of an X-class solar flare on 14 May 2013. 
In Sect. 2, we briefly present the observations and power spectral analysis 
methods for detecting QPPs. Sect. 3 describes the results of the QPPs 
observed in different energy bands of the flaring loop. Exploration of 
various harmonics of appropriate MHD modes leaving their imprints on 
selected SXR while excited in the flaring loop system is discussed in 
Sect. 4. In Sect. 5, MHD seismology of flaring plasma using the observed 
QPPs are discussed along with conclusions.

\section{Observations and Analysis Methods}

The X 3.2 class solar flare on 14 May 2013 is investigated in the present study. 
It started at 00:42:39 UT, reached its maximum at 01:07:41 UT, gradually 
decayed and ended at 03:14:40 UT in GOES SXR observations. Its coordinate 
at the Sun is N11E74 in an active region NOAA 11748.
The flare region is detected by both RHESSI and Nobeyama Radioheliograph (NoRH) 
in an integrated mode providing flaring light curves in its decay phase in X-rays and radio 
respectively. Figures 1 (a,b,c) represent some features of this solar 
flare and related emissions.

Figure ~1 (a) shows time profiles of the flare in soft X-rays taken with ~GOES~ satellite 1.0--8.0 \AA~ channel and microwave emission during this flare at 17 and 34~ GHz observed by  NoRH. NoRH provides intensity (Stokes parameter I=R+L) and circular polarization (Stokes parameter V=R-L) 
images at 17 GHz with a temporal resolution of 1.0 sec (Nakajima et al. 1994).
The fluxes (Figure 1(a), middle panel) and correlation curves (Figure 1(a), bottom panel) at 17 GHz and 34 GHz obtained from NoRH vary with the same pattern during the flare process. However, for the flare peak, 
the 17 GHz flux is larger than the 34 GHz and afterwards both fluxes become 
comparable with each other. The radio emission is probably generated by 
gyro-synchrotron motion of accelerated electrons during the flare peak phase, 
while in the long decay phase the thermal free–free emissions dominate.
The full Sun image obtained from NoRH at 17 GHz (Stokes parameter I=R+L) 
and the corresponding flaring region of the day under study is shown in Figure 1 (b).

Figure 1(c) shows this localized temporal span of the light curve during the decay 
phase of the flare detected by RHESSI. In these data sets of 3--6 KeV, 6--12 KeV 
and 12--25 KeV energy bands, best-fit exponential functions of the form $~I=I_0 \times e^{-bt}$ are fitted for the time period  between 01:23:40 UT and  01:51:56 UT to remove the long-term
background flare variations and  corresponding graphs are shown in Figure 2. For 3--6 KeV energy band, the values $I_0$ and $b$ are 349.9 and -0.003996; for 6--12 KeV 
data set, these values are 4265 and -0.004179; and for 12--25 KeV energy band, the corresponding values are 1960 and  -0.005061 respectively.

We have studied the QPPs present in the de-trended signal during the decay phase of the flare by 
conventional Fast Fourier Technique (FFT), Maximum Entropy Method (MEM) and Lomb-Scargle (LS) 
periodogram analysis method. The MEM is an alternative method to FFT which avoids the limited 
resolution and power $'$leaking$'$, due to the windowing of data, present in the FFT, and it belongs 
to the class of methods which fit a satisfied model to the data. The parameters of a maximum 
entropy spectral estimation are equivalent to the $'$ones$'$' in the auto-regressive (AR) model 
of a random process in real domain \citep{Burg}.

On the other hand, the Scargle periodogram \citep{Sca1982,Horne1986} 
is an important algorithm for time series analysis of unevenly sampled data. Being quite 
powerful for finding and testing, the significance of weak periodic signals through false 
alarm-probability (FAP), a simple estimate of the significance of the height of a peak in 
power spectrum can be derived. In this method, the confidence levels have been calculated 
according to the recipe given in \cite{Horne1986}. In case of FFT and MEM spectral 
decomposition techniques, we have calculated the mean ($\mu$) and standard deviation ($\sigma$) of the 
data points. We have then calculated the confidence levels as $\mu$ + 2$\sigma$ = 95.4 \%; $\mu$ + 3$\sigma$ = 99.7 \% 
and $\mu$ + 4$\sigma$ = 99.99 \% respectively. X-axis of FFT, MEM and LS graphs are drawn on a log scale.

In the present study, the periods which are common in all the methods having power $>$ 95 \% 
confidence levels are considered only for final discussion and calculation. In order to 
justify the validity of the peaks found from the FFT, MEM and LS methods, the confidence 
limits for different peaks obtained from the power spectra of  different energy bands have been 
calculated \citep{Haber1969,Chowdhury2006,Chowdhury2009}. In every case, 
confidence levels are above 99\%. In effect, we attempt to determine the interval in which any 
hypothesis concerning the periodicity 
of a certain solar event might be considered tenable and outside which any hypothesis would be 
considered untenable. The confidence limits (CL) are evaluated by generating a sample of 100 data 
points equally on both sides of a particular peak. This method is repeated for all the peaks under 
consideration within a spectrum. The peak that is sharp gives the minimum value of standard error 
and closer values of confidence limits.

The confidence interval provides the lower and upper limits to which the population parameter has a high
 probability of being included. The population parameter standard deviation $\sigma$
can be calculated from the following formula:

\begin{equation}
\sigma =\{\Sigma{T_i^2}/(N-1)-(\Sigma{T_i^2/N(N-1)\}}^{0.5}\,\,\,\,\,\,\,\,\,\,\,\,\,\,\,\,\,\,\,\,\,\,\,\,\,\,\,\,\,\,\,\,\,\,\,\,\,\,\,\,\,
\label{eq1}
\end{equation}

Here ${T_i}$ represents the ${i}$ value of the data under study and N is the total number of data points.  
The standard error $\mathrm{SE}_{\mathrm{m}}$ is the standard deviation of the sample mean (from 
sampling distribution) which is estimated as: 

\begin{equation}
SE_m={\sigma/(N)}^{0.5} \,\,\,\,\,\,\,\,\,\,\,\,\,\,\,
\label{eq2}
\end{equation}

The confidence limits (CL) for 99\% confidence can be computed as 

\begin{equation}
CL=T_{ave}\pm2.58(SE_m),\,\,\,\,\,\,\,\,\,\,\,\,\,\,\,\,
\label{eq3}
\end{equation}
where ~${T_{ave}}$~ is the mean value of the time period of the sample data points.

To determine confidence limits we have used the formula of 
${Z_a/2} \times \sigma\sqrt{(n)}$. Here ${Z_a/2}$ is the confidence coefficient, a is confidence level, $\sigma$ is standard deviation, 
and n is sample size. Then, we have converted the 99 \% to a decimal 0.99 and divide it by 2 to get 0.495.  According to the table the closest value corresponding 
to 0.495 is 2.60. Then we have multiplied 2.6 by 0.99 (critical value by our standard error) 
and we get 2.574 ($\sim$2.58) and this is our margin of error.

We have then calculated the cross-correlation coefficients (CC) between de-trended data 
of three different energy channels observed by RHESSI to study the time-delay with various 
time-lags (0, $\pm$4, $\pm$8...$\pm$ 50 seconds). Each of these data sets has been shifted forward 
or backward and the cross-correlation coefficient has been calculated after shifting the 
time-lag (L) corresponding to the best correlation of the time phase under investigation. 
The results are shown in Figures 6 (a,b,c). 

\section{Results}
\subsection{Periodogram Analysis}
Figure 3(a, b, c) represent the FFT, MEM and Scargle spectrum of 3--6 KeV RHESSI 
channel and the corresponding results are given in Table 1.

Figures 4 (a, b, c) represent the FFT, MEM and Scargle spectrum of 6-12 KeV RHESSI data and the corresponding results are given in Table 2.

Figures 5 (a, b, c) represent the FFT, MEM and Scargle spectrum of 12-25 KeV RHESI data and the corresponding results are given in Table 3.
Thus we find, in all methods and in all energy bands, the common periods to be $\sim$71 sec (P$_{1}$) and $\sim$53 sec (P$_{2}$). However, for the energy band of 12--25 KeV, an additional period $\sim$39 sec (P$_{3}$) is found in all spectral decomposition techniques.  The period ratios are estimated as P$_{1}$/P$_{2}$ = 71/ 53 $\sim$ 1.339; P$_{1}$/2P$_{2}$$\sim$0.669; P$_{1}$/P$_{3}$=71/39$\sim$1.82 or P$_{3}$/P$_{1}$$\sim$0.549.

\begin{table}
\caption{List of periods detected with their Standard Errors for Confidence Limits in the decay phase of solar flare in energy band 3 –- 6 KeV}
\label{TabIndices}
\centerline{\begin{tabular}{lccccccc}
\hline
\hline
Energy band & & & Identified spectral peaks (sec) \\
&  & 1 & 2 & 3 \\
\hline
\multicolumn{6}{c}{\it {\bf (a) By FFT}}\\
3-6 KeV & $\mathrm{T}_{\mathrm{ave}}$ & 71.56 & 53.34 & 38.62 \\
& $\mathrm{SE}_{\mathrm{m}}$ & 0.035 & 0.0195 & 0.0101 \\
& T & 71.56 $\pm$ 0.0903 & 53.34 $\pm$ 0.0503 & 38.62 $\pm$ 0.0261 \\
\multicolumn{6}{c}{\it {\bf (b) By MEM}}\\
& $\mathrm{T}_{\mathrm{ave}}$  & 53.36 & 71.74 & 32.71 \\
& $\mathrm{SE}_{\mathrm{m}}$ & 0.274 & 0.457 & 0.102 \\ 
& T & 53.36 $\pm$ 0.7069 & 71.74 $\pm$ 1.1791 & 32.71 $\pm$ 0.2632 \\
\multicolumn{6}{c}{\it {\bf (c) By Scargle Technique}}\\
& $\mathrm{T}_{\mathrm{ave}}$  & 71.89 & 53.47 & 38.80 \\
& $\mathrm{SE}_{\mathrm{m}}$ & 0.0352 & 0.0192 & 0.0101 \\
& T & 71.89 $\pm$ 0.0908 & 53.47 $\pm$ 0.0495 & 38.80 $\pm$ 0.0260& \\
\hline
\end{tabular}}
\end{table}

\begin{table}
\caption{List of periods detected with their Standard Errors for Confidence Limits in the decay phase of solar flare in energy band 6 –- 12 KeV}
\label{TabIndices}
\centerline{\begin{tabular}{lccccc}
\hline
\hline
Energy band & &  Identified spectral peaks (sec) \\
&  & 1 & 2   \\
\hline
\multicolumn{5}{c}{\it {\bf (a) By FFT}}\\
6-12 KeV & $\mathrm{T}_{\mathrm{ave}}$ & 71.96 & 53.34 \\
& $\mathrm{SE}_{\mathrm{m}}$ & 0.0363 & 0.0203 \\
& T & 71.96 $\pm$ 0.0937 & 53.34 $\pm$ 0.0524  \\
\multicolumn{5}{c}{\it {\bf (b) By MEM}}\\
& $\mathrm{T}_{\mathrm{ave}}$  & 71.80 & 53.43 \\
& $\mathrm{SE}_{\mathrm{m}}$ & 0.479 & 0.2803 \\ 
& T & 71.80 $\pm$ 1.2358 & 53.43 $\pm$ 0.7232  \\
\multicolumn{5}{c}{\it {\bf (c) By Scargle Technique}}\\
& $\mathrm{T}_{\mathrm{ave}}$  & 71.73 & 53.41 \\
& $\mathrm{SE}_{\mathrm{m}}$ & 0.0354 & 0.0183 \\
& T & 71.73 $\pm$ 0.0913 & 53.41 $\pm$ 0.0472 & & \\
\hline
\end{tabular}}
\end{table}

\begin{table}
\caption{List of periods detected with their Standard Errors for Confidence Limits in the decay phase of solar flare in energy band 12 -- 25 KeV}
\label{TabIndices}
\centerline{\begin{tabular}{lccccc}
\hline
\hline
Energy band & & & Identified spectral peaks (sec) \\
&  & 1 & 2 & 3 & 4 \\
\hline
\multicolumn{6}{c}{\it {\bf (a) By FFT}}\\
12-25 KeV & $\mathrm{T}_{\mathrm{ave}}$ & 71.08 & 53.14 & 38.88 \\
& $\mathrm{SE}_{\mathrm{m}}$ & 0.0312 & 0.0185 & 0.0106 \\
& T & 71.08 $\pm$ 0.0903 & 53.34 $\pm$ 0.0503 & 38.88 $\pm$ 0.0273 \\
\multicolumn{6}{c}{\it {\bf (b) By MEM}}\\
& $\mathrm{T}_{\mathrm{ave}}$  & 71.52 & 53.36 & 32.69 & 39.76\\
& $\mathrm{SE}_{\mathrm{m}}$ & 0.537 & 0.278 & 0.104 & 0.213\\ 
& T & 71.52 $\pm$ 1.385 & 53.36 $\pm$ 0.7172 & 32.69 $\pm$ 0.2683 & 39.76 $\pm$ 0.5495 \\
\multicolumn{6}{c}{\it {\bf (c) By Scargle Technique}}\\
& $\mathrm{T}_{\mathrm{ave}}$  & 72.15 & 53.60 & 38.78 \\
& $\mathrm{SE}_{\mathrm{m}}$ & 0.0367 & 0.018 & 0.0104 \\
& T & 72.15 $\pm$ 0.0947 & 53.60 $\pm$ 0.0464 & 38.78 $\pm$ 0.0268& \\
\hline
\end{tabular}}
\end{table}

\subsection{Cross-correlation Analysis}

We also study lagging and leading behaviors of RHESSI 3--6 KeV, 6--12 KeV and 12--25 KeV channels during the decay phase of the flare by the cross-correlation technique and the plots are displayed in Figures 6(a, b, c). It is detected that there is no lag/lead between different energy channels demonstrating very high correlation of the signals in all observational bands.
From figure 6 (a), it is detected that maximum cross-correlation coefficient is 0.68$\pm$0.0168 at 0 second.  Figure 6 (b) exhibits that maximum cross-correlation coefficient is 0.643$\pm$0.0195 at 0 second.  The maximum cross-correlation coefficient is 0.781$\pm$0.0264 at 0 second between the energy range 6--12 KeV and 12--25 KeV (Fig. 6c).

\subsection{Morphology of the Flaring Loop System}

GOES SXI image of the active region and its loop system beneath where the flare occurred is shown in Fig. 7 (right panel), while the same active region in 17 GHz NoRH observations is also displayed in Fig. 7 (left panel). The North-South 
axis in the GOES map is inclined for aproximately 20 degrees clockwise relatively to the vertical direction. Therefore
NoRH map shows the true position of the AR while GOES SXI image shows the rolled partial view. 
The soft X-ray enhancement occurred in the loop system during the evolution of the solar flare. The 
length of the flaring loop system is derived as L$\sim$190 Mm while its width is $\sim$12 Mm. The 
magnetic configuration of the flaring site obtained for 01:00 UT when approximate loop-length and width are derived,
remains almost the same and changes insignificantly in 20 min later when QPOs evolve in the X-ray light curves. The 
loop-length is estimated by tracking the pixels along the loop. However, it should be noted that the 
estimated length is the lower bound projected length. The loop width is also estimated across the loop 
near its apex. 

\section{Results and Discussion}

The observed periods are most likely associated with the first two harmonics of the fast 
magnetoacoustic wave modes in the flaring loop system. The phase speed of the fundamental mode of 
magnetoacoustic wave is $V_{ph}$ = 2$L/P $$\sim$ 5200 km/s, which seems to lie with the fastest magnetoacoustic branch (tubular-mode) within the loop. Since the QPPs modulate the emissions coming 
from the flaring loop system, they are compressible fast modes known as fast magnetoacoustic sausage modes.

The dispersion relation of fast-sausage waves in straight cylindrical tube includes both trapped and leaky modes depending on the plasma and magnetic field parameters \citep{Edwin1983}. 

The cut-off wavenumber k$_{c}$ is written as given by \cite{Edwin1983} and \cite{Rob1984}:

\begin{equation}
k=k_c={\left[{\frac{V_{A}^2}{V_{Ae}^2-V_{A}^2}}\right]}^{1/2}{\frac{j_{0}}{a}},
\label{eq4}
\end{equation}

where V$_{A}$ and V$_{Ae}$ are Alfv\'en speeds inside and outside the loop, a is the loop radius and j$_{0}$= 2.4 is 
the first zero of the Bessel function. The modes with k $>$ k$_{c}$ are trapped in the loop system, while 
the modes with k $<$ k$_{c}$ are leaky. The trapped fundamental global sausage mode (P = 72 s) can be realized 
in this loop system only if the density contrast is:

\begin{equation}
{\frac{L}{2a}}<{\frac{\pi V_{Ae}}{2j_{0}V_{Ao}}}\approx0.65{\sqrt{\frac{\rho_{o}}{\rho_{e}}}},
\label{eq5}
\end{equation}

which is equal to 593 for a given loop length L$\sim$190 Mm and width a$\sim$6 Mm.  Therefore, the observed flaring loop must be very thick loop with significant density contrast to possess the global sausage modes. The maximum phase speed of the sausage wave 5200 km/s is considered as an external Alfv\'en speed (V$_{Ae}$), while V$_{Ao}$ is the internal Alfv\'en speed within the denser loop system.  Using Eq. (2), we can estimate the ratio of external and internal Alfv\'en speed (V$_{Ae}$/V$_{Ao}$) in the loop system as $\sim$24.0. The phase speed of the fundamental mode sausage wave (V$_{ph}$) can be maximum limit to the external Alfv\'en speed, i.e., V$_{Ae}$=V$_{ph}$$\sim$5200 km/s. Therefore, the internal Alfv\'en speed (V$_{Ao}$) within the loop will be 216 km/s. The diagnosed internal Alfv\'en speed (V$_{Ao}$) of the flaring loop system is 216 km/s. The Alfv\'en speed inside the bulky flaring loop is significantly less compared to that in the outside ambient medium. This is expected as the density of such loops is higher compared to the normal loops as well as ambient plasma. This loop system does not possess the global sausage modes with period greater than PGSM$\sim$2$\pi$/k$_{c}$ V$_{Ae}$$\sim$ 2L/V$_{Ae}$$\sim$73.0 s \citep{Pas2007}. It should be noted that L=$\pi$/ k$_{c}$. The longest period of the trapped sausage mode can have the period of $\sim$73 s. The observed period of fundamental mode sausage waves is less than this maximum period. Therefore, this period and its harmonics are well accepted in the given morphology and density contrast of the loop system.

Further, we have also found zero time lag between different soft X-ray energy bands. This zero lag in all bands depicts the periodicity that the wave concurred by sausage mode is not a propagating wave, but it is standing wave in nature. Therefore, we basically observe fast MHD pulsation of short periods excited due to flare activity in the flaring loop system. As a result, the observed denser loops can support the non-leaky, trapped global sausage mode in the estimated length and width as well as density contrast \citep{Nak2003,Asc2004,Srivastava2008}. Such flaring/post-flare loops usually have a very high density contrast of the order of 10$^{2}$--10$^{3}$. The density evolves very rapidly during the flare heating and makes such loops highly denser. The observed loops are the X-class flaring loops where large heating is likely to take place during the strong X-class flare. Therefore, the estimated high density contrast is likely to occur and also falls in the expected range. Thus, the trapped sausage mode is probably excited in the loop system. 

The ratio between the periods of the fundamental and the first harmonics of sausage waves is equal to P$_{1}$/P$_{2}$$\sim$1.34, which is shifted from 2.0. Similar physical behavior is observed for the fast-kink oscillations \citep{Ver2004,Tom2007,De2007,White2012,Sri2013}, for sausage waves \citep{Srivastava2008}, and for slow waves \citep{Srivastava2010}. The deviation of P$_{1}$/P$_{2}$ from 2.0 in homogeneous loops is very small due to the wave dispersion \citep{Mc2006,Andries2005}. However, longitudinal density stratification causes significant shift of P$_{1}$/P$_{2}$ from 2.0 \citep[and references therein]{Andries2005,Andries2009,Mc2006,Tom2007,Mc2010,Mac2011,Luna2012}. The longitudinal variation of density in a coronal loop is the most likely cause for the departure of the observed period ratio of the sausage waves. However, magnetic field geometry may also play an important role in its deviation \citep{Mac2011}. \cite{Mac2011} have shown in their analytical calculations that as V$_{Ae}$/V$_{Ao}$ tends to infinity the period ratio of the sausage waves (P$_{1}$/2P$_{2}$) approaching towards 0.5 in their slab-model. Normally, it should be equal to 1.0. Taking account of our observed parameters, we obtain a very high ($\sim$24) external to internal Alfv\'en speed ratio. The period ratio (P$_{1}$/2P$_{2}$) is significantly less, i.e., 0.65. This observation is qualitatively in agreement with the analytical solution of \cite{Mac2011} that period ratio reduces significantly for higher Alfv\'en velocity ratios.  

 In the present work, we find that QPPs are most likely caused by the sausage modes, and therefore, based on their signature local plasma properties are diagnosed. However, there may be other possible mechanisms which may be potentially responsible for QPP modulation. A few examples may be cited as the periodic variations of the angle between the line-of-sight and magnetic field by weakly compressible kink waves \citep{Coop2003}, or by excitation of QPPs by MHD oscillation of an external loop \citep{Nak2009}.  Recently, \cite{Kol2015} have analyzed the multiple MHD modes during the impulsive phase of the same flare that was approximately 18 min before the time interval analyzed in our present paper and lasting for a few minutes only. The signatures of waves are revealed with the mean periods of 15, 45, and 100 s. The 15 s and 100 s period are associated with the MHD modes rapidly decaying within a few wave periods. They are interpreted as leaky sausage modes and the kink modes respectively. The 45 second periodicity showed rather non decaying behavior and the period similar to the findings of our present paper. This example shows that in the flaring region, the plasma and magnetic field conditions are rapidly varying. Therefore, even a single flare energy release over different epoch in the given duration of the flare may give rise to the evolution of different MHD modes and their properties. These properties are highly linked to the local plasma and magnetic field conditions, and can be used in diagnosing highly variable atmosphere there. 

\medskip
{\bf Acknowledgments}
We are grateful to the anonymous reviewers for their valuable comments. P. Chowdhury acknowledges BK21 plus program  of the National Research Foundation (NRF) funded by the Ministry of Education of Korea. The research R. Sych has been funded by Chinese Academy of Sciences President's International Fellowship Initiative, Grant No. 2015VMA014 and by a Russian Foundation for Basic Research, Grants 13-02-00044 and 14-02-91157.
\medskip


\begin{figure*}
\begin{center}
\includegraphics[width=13.0 cm]{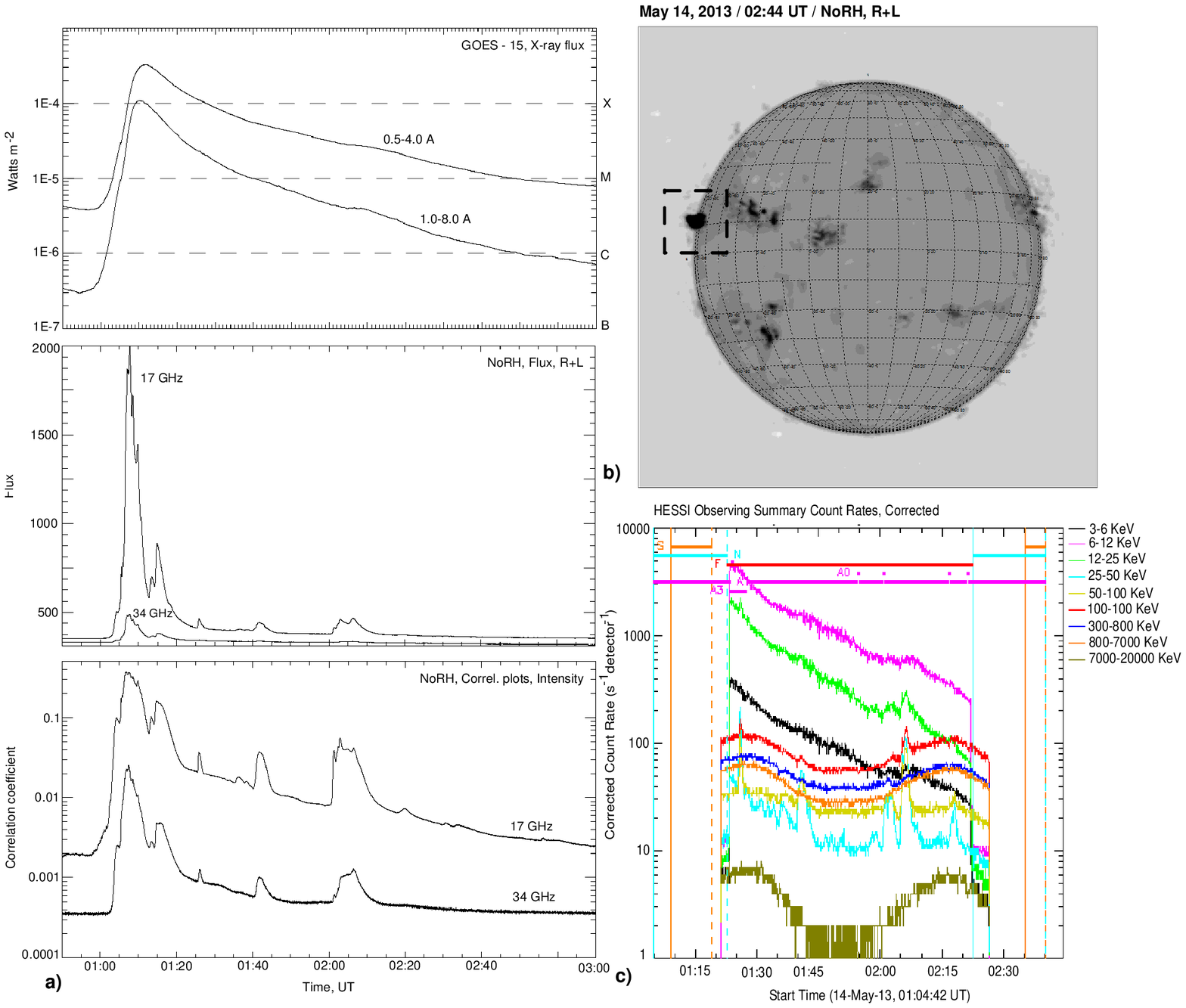}
\end{center}
\caption{(a) GOES plot showing temporal evolution of soft X-rays (top panel). The microwave emission measured by NoRH at 17 and 34 GHz - integral flux (middle panel) and correlation plots (bottom panel). (b) Full Sun in intensity NoRH channel at 17 GHz.
The flare region is indicated by the solid box. (c) Localized temporal span of the light curves of different energy bands during
the decay phase of the flare detected by RHESSI on May 14, 2013}
\label{Fig1}
\end{figure*}

\begin{figure*}
\begin{center}
\includegraphics[width=12.0 cm]{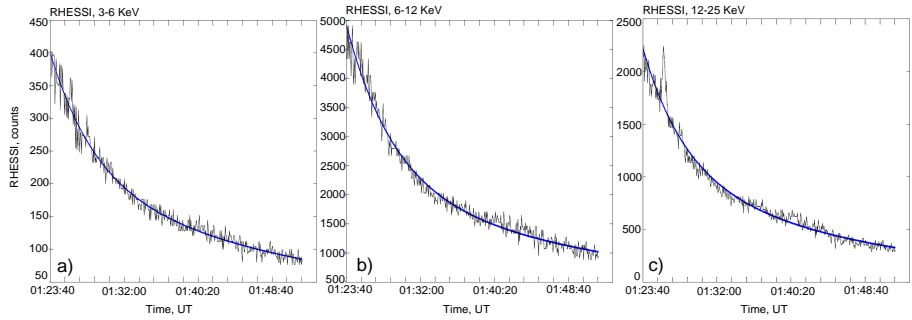}
\end{center}
\caption{RHESSI soft X-ray light curves on May 14, 2013, from 01:23:40 to 01:51:56 UT. (a) Blue line is the trend fitting X-ray light curve for 3--6 Kev data; (b) same for energy band
6--12 keV and (c) same for energy band 12--25 Kev.}
\label{Fig2}
\end{figure*}

\begin{figure*}
\begin{center}
\includegraphics[width=12.0 cm]{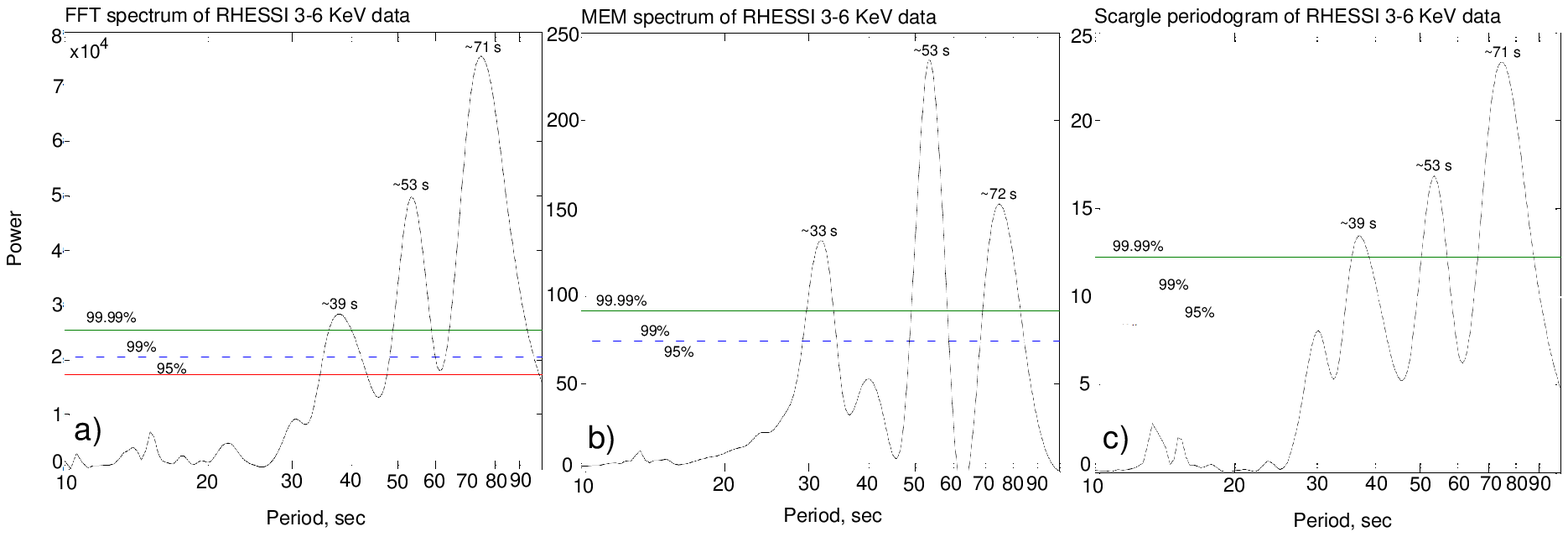}
\end{center}
\caption{(a) FFT spectra of RHESSI 3--6 Kev data; (b) MEM spectra for the same data
and (c) Scargle spectra of the same data.}
\label{Fig3}
\end{figure*}

\begin{figure*}
\begin{center}
\includegraphics[width=12.0 cm]{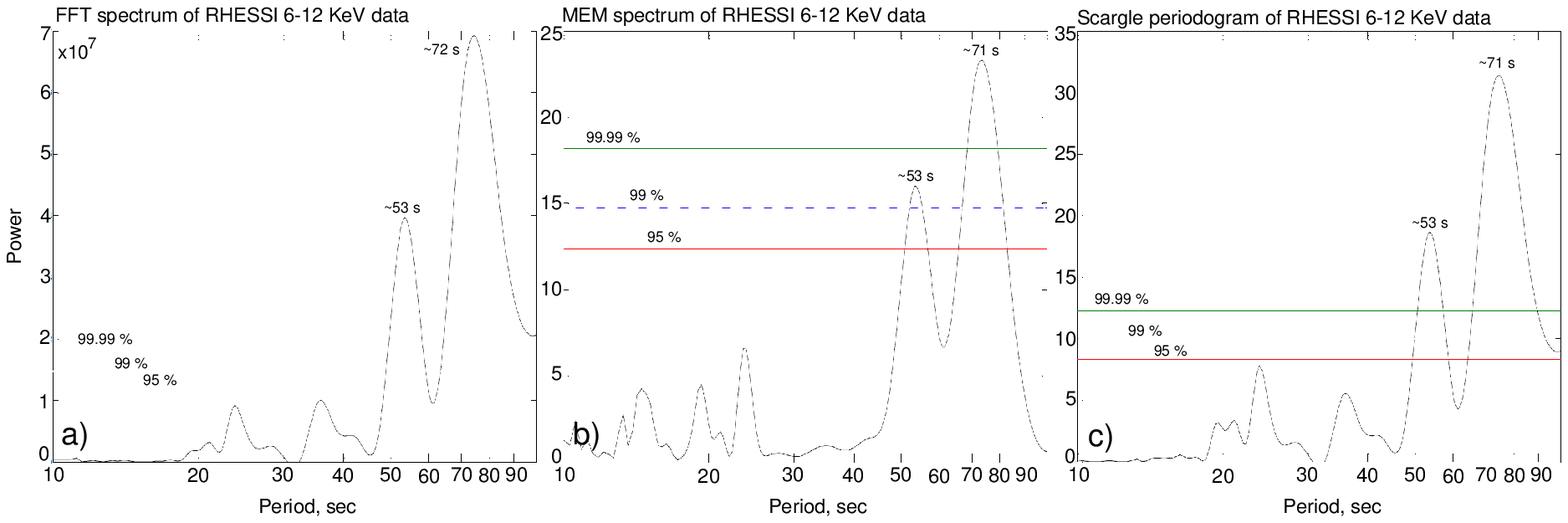}
\end{center}
\caption{(a) FFT spectra of RHESSI 6--12 Kev data; (b) MEM spectra for the same data
and (c) Scargle spectra of the same data.}
\label{Fig4}
\end{figure*}

\begin{figure*}
\begin{center}
\includegraphics[width=12.0 cm]{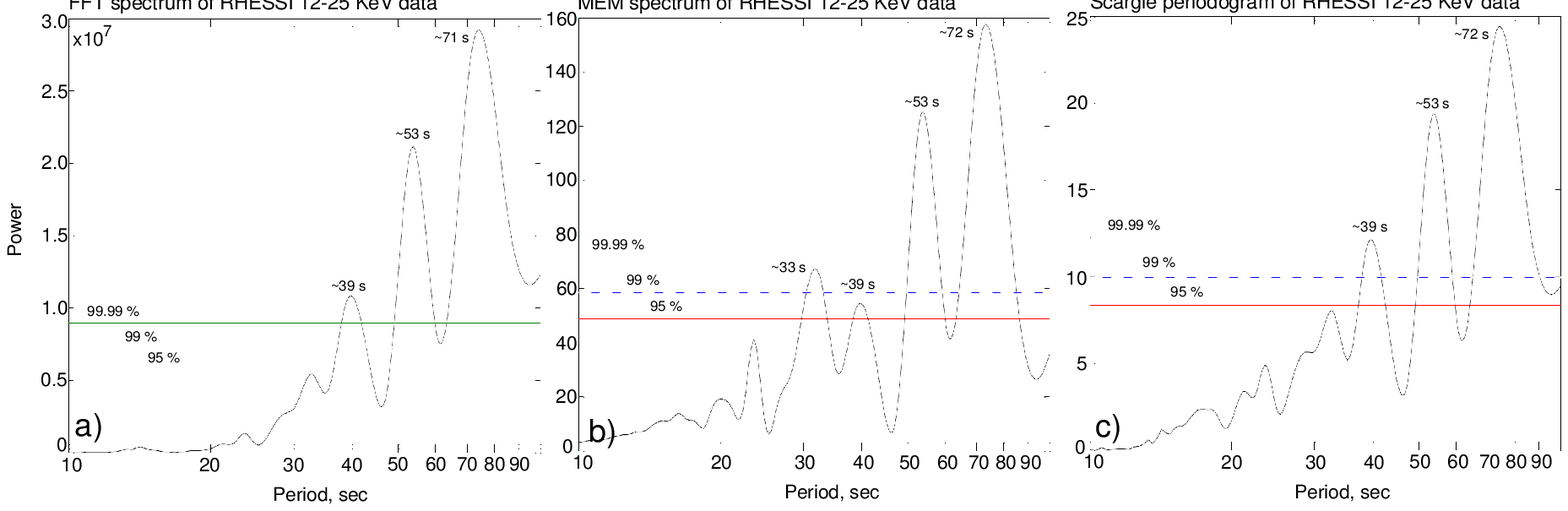}
\end{center}
\caption{(a) FFT spectra of RHESSI 12--25 Kev data; (b) MEM spectra for the same data
and (c) Scargle spectra of the same data.}
\label{Fig5}
\end{figure*}

\begin{figure*}
\begin{center}
\includegraphics[width=12.0 cm]{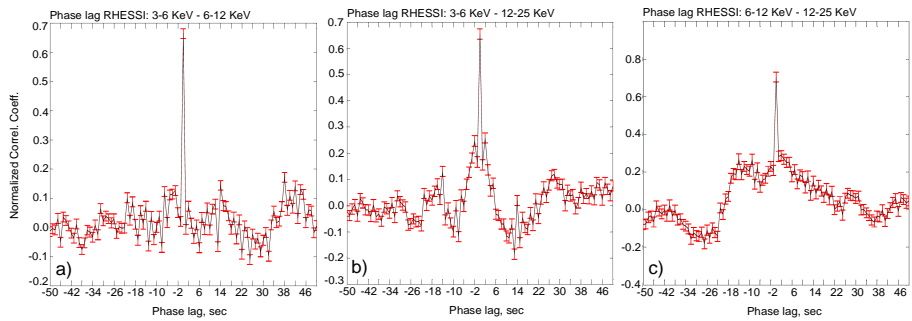}
\end{center}
\caption{(a) Cross-correlation analysis between RHESSI 3 --6 and 6--12 Kev data sets ;
(b) Cross-correlation analysis between RHESSI 3--6 and 12--25 Kev data ; (c) Cross-correlation analysis between RHESSI 6--12 and 12--25 Kev data set.}
\label{Fig6}
\end{figure*}

\begin{figure}
\centerline{
\includegraphics[width=12.0cm]{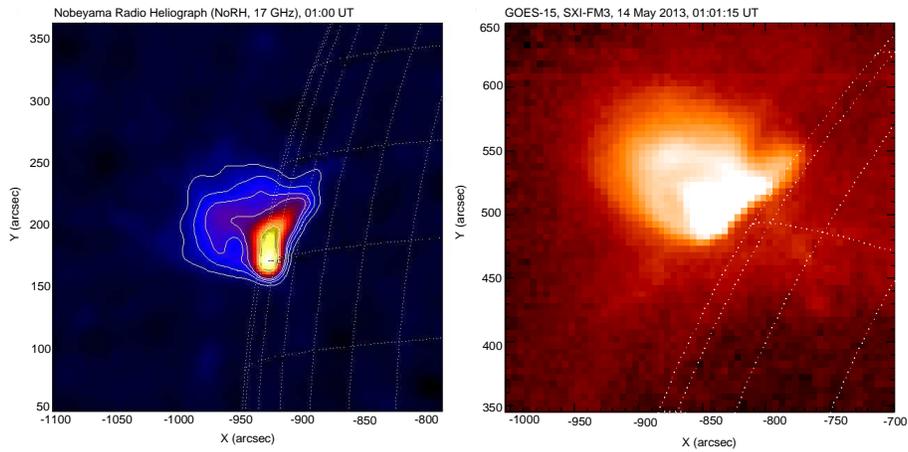}}
\caption{Left-panel : NoRH image showing the position of flaring active
region. Right-panel : Partial view of the GOES SXI image of X-class flaring region (negative
color image) and loop system beneath which the flare occurred. The soft X-ray enhancement
occurred in the loop system is shown during the evolution of the solar flare. The North-South 
axis in the GOES map is inclined for aproximately 20 degrees clockwise relatively to the vertical direction. Therefore
NoRH map shows the true position of the AR while GOES SXI image shows the rolled partial view.}
\label{Fig7}
\end{figure}

\end{document}